# The refined EUV mask model

I.A. MAKHOTKIN[1*], M. WU[1], V. SOLTWISCH[2], F. SCHOLZE[2], V. PHILIPSEN[1]

[1.]*IMEC, Kapeldreef 75, 3001 Leuven, Belgium*

[2.] *Physikalisch-Technische Bundesanstalt, Abbestr. 2-12, 10587 Berlin, Germany*

**Abstract.**

A refined model of an extreme ultraviolet (EUV) mask stack consisting of the Mo/Si multilayer coated by a Ru protective layer and a TaBN/TaBO absorber layer was developed to facilitate accurate simulations of EUV mask performance for high-NA EUV photo-lithography (EUVL) imaging. The model is derived by combined analysis of the measured EUV and X-ray reflectivity of a state-of-the-art mask blank. These two sets of measurements were analyzed using a combined free-form analysis procedure that delivers high-resolution X-ray and EUV optical constant depth profiles based on self-adapted sets of sublayers as thin as 0.25nm providing a more accurate description of the reflectivity than obtained from only EUV reflectivity. "Free-form analysis" means that the shape of the layer-interfaces in the model is determined experimentally and is not given a priori by the structure model. To reduce the numerical effort for EUV imaging simulations a low-resolution model of the multilayer and absorber stack with sublayer thicknesses larger than 2nm, that fits to only the EUV reflectance, was derived from the high-resolution model. Rigorous high-NA EUVL simulations were done to compare the performance of the new model to our previous work [1].

*Corresponding author

*i.makhotkin@utwente.nl (Igor A. Makhotkin)*

## 1. Introduction

The EUV mask is one of the key components of the photo lithography setup that can be designed and optimized by the end-user. The mask design largely determines the EUV lithography performance. The development of new masks is based on the large-scale numerical optimization of the EUV imaging quality by varying the parameters of the absorber patterns. Such mask optimization, as for example discussed in [2], can be used for the mitigation of mask 3D effects in order to improve EUV imaging.

A comprehensive understanding of the EUV mask stack (multilayer and absorber) is required to explore EUV imaging at high NA using rigorous mask 3D lithography simulations and to support EUVL at current NA 0.33 using full-field design modeling software. Current mask model was presented in 2013 [1] and is calibrated to the EUV reflectivity measured from the, at that time, state-of-the-art mask blank. The recent developments in mask making process as well as in EUV multilayer metrology calls for the model update.

The detailed study of the periodic multilayer EUV reflectivity (EUVR) analysis [3] showed that the single-wavelength EUV-only reflectivity measurement, although being very sensitive to minor structural changes of the multilayer, generally cannot be used for accurate determination of the sample structure because of the highly correlated influence of multilayer structural parameters, for example multilayer thickness ratio, densities and stoichiometries of layer materials. A minor change of the ratio between layer thicknesses in the multilayer Comodel will change the simulated EUVR curve, however this change can be compensated by the change of total bi-layer thickness and layer density, making it impossible to determine accurate structural parameters from EUVR fit only [3].

To solve such correlation a combination of X-ray reflectivity (XRR) and EUVR measurements [3] was proposed, as well as the more complex combination of EUVR with scattering and X-ray fluorescence measurements [4]. In this paper we will present a mask stack model – that simultaneously fits to both EUVR and XRR measurements of the actual mask. This combination allows us to accurately describe the structure and optical properties of the mask stack for EUV imaging. The implementation of the presented mask stack model in mask 3D aware simulation tools will enhance their predictive and pre-compensation power.

## 2.   Mask measurements and data analysis

The presented model is built based on analysis of two sample structures from current state-of-the-art EUV mask blanks. The first sample (referred to as MLM) is a periodic Mo/Si multilayer mirror containing 40 Mo-Si bilayers with the period thickness of 7nm deposited on $SiO_2$ glass substrate. The multilayer is covered with a 3nm protective Ru layer. The second sample, referred to as absorber, is TaBN-TaBO absorber bi-layer with a thickness of 58nm and 2nm respectively deposited on the glass substrate coated with a 3nm thin Ru layer. In this set both samples have identical Ru layers and therefore the obtained models can be merged by overlapping the identical Ru layer.

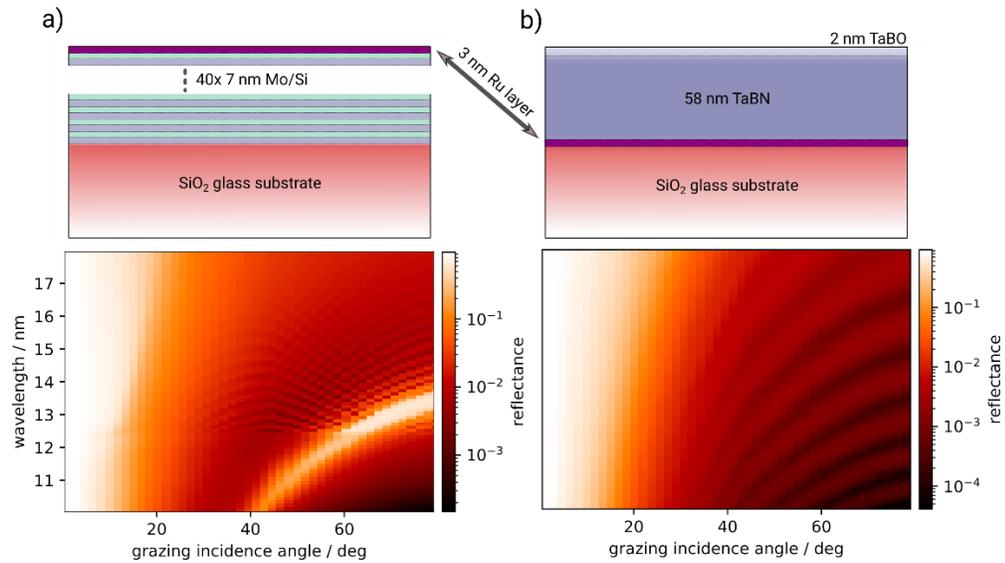

Figure 1. Scheme of the a) Mo/Si multilayer and the b) TaBN/TaBO absorber layer with corresponding measured EUV reflectivity maps.

The grazing incidence XRR is measured using Malvern Panalytical X'pert MRD XL multipurpose diffractometer equipped with a Cu X-ray source. For XRR measurements the Cu K X-ray beam was conditioned, using the combination of parallel beam mirror and 4x Ge 220 monochromator enabling measurements over the large dynamic range of $10^7$ counts per second with high resolution of 0.012 degrees, determined by the beam divergence. The EUV reflectivity measurements were conducted at the soft X-ray beamline of the Physikalisch-Technische Bundesanstalt at the electron storage ring BESSY II, which covers the photon energy range from 50 eV to 1900 eV. The beamline is equipped with a ultra-high vacuum lubrication free Ellipso-Scatterometer [5, 6]. The EUV reflectivity curves were measured using s-polarized radiation for the full accessible angular range from grazing incidence to near normal and mapped over the incidence wavelength from 10nm to 18nm. The experimental uncertainty is considered for every angle of incidence scan separately to take into account for beam instabilities during the mapping. The schemes of the two measured samples together with EUV reflectivity maps are presented in Fig. 1.

For the analysis of the sample structures, only the incidence angle scan at 13.5nm wavelength was used from the measured EUVR map. The data used for the analysis of XRR and EUVR are shown in Fig. 2a, 2b and 3a, 3b.

The analysis was performed by following steps. The high-resolution profile of the MLM and absorber. samples is obtained using a modified free-form approach similar to discussed by Zameshin et.al. in [7]. For a free-form analysis of X-ray and EUV reflectivity curves, the analyzed film is modeled as a set of thin sub-layers where the optical constants of each sublayer are determined by sub-layer chemical composition and density [3]. The stoichiometry values and densities were coded using the array of integers $P$ as described in [8]. These values of $P$ and total layer thicknesses were the only fitting parameter. In this way, a set of consistent optical constant profiles can be calculated for various wavelengths having equal sub-layer densities and stoichiometries and changing only wavelength-dependent atomic scattering factors [9]. The extended description of applied data analysis procedure is beyond the scope of current paper and therefore will be published elsewhere. The maximum thickness of the sublayer defines the in-depth resolution of the optically constant profile of the thin film and is determined by the measurement with the highest resolution. For both samples the X-ray reflectivity determined the resolution of the high-resolution model to be 0.25nm corresponding to the measurement range of 9 degrees (Fig. 2a) at 0.154nm wavelength (see [7] for more details). For the XRR analysis of the absorber sample a sub-layer thickness of 0.5nm was used as the XRR measurement is informative only till 4 degrees (Fig. 3a). The profile steps in the Ru layer were kept as small as in the MLM model for the ease of the later merger of the models. Following the same logic, the minimal steps in both profiles to fit EUV reflectivity can be as large as 3.3nm assuming measurements at 13.5nm till 88 degrees grazing (2 degrees normal) incidence. Regardless of their low in-depth resolution, the EUVR data contribute to the combined analytical accuracy of the determination of densities and chemical stoichiometries of absorber and MLM models[3].

However, the fact that the optical constant model builds up from sublayers with thickness of ~3.3nm can be accurate enough to fit the measured EUVR data, means that for EUV lithography simulations a low-resolution model can be build based on EUVR-only fitting, that produce as accurate as high-resolution model simulation results. Consequently, as a second step of the analysis we obtain here the low-resolution models for MLM and absorber layer by fitting only the EUVR data. The initial guess model here was build based on the high-resolution model by combining its thin sublayers to thicker ones with averaged stoichiometries and densities.

## 2.1 The Mo/Si multilayer model

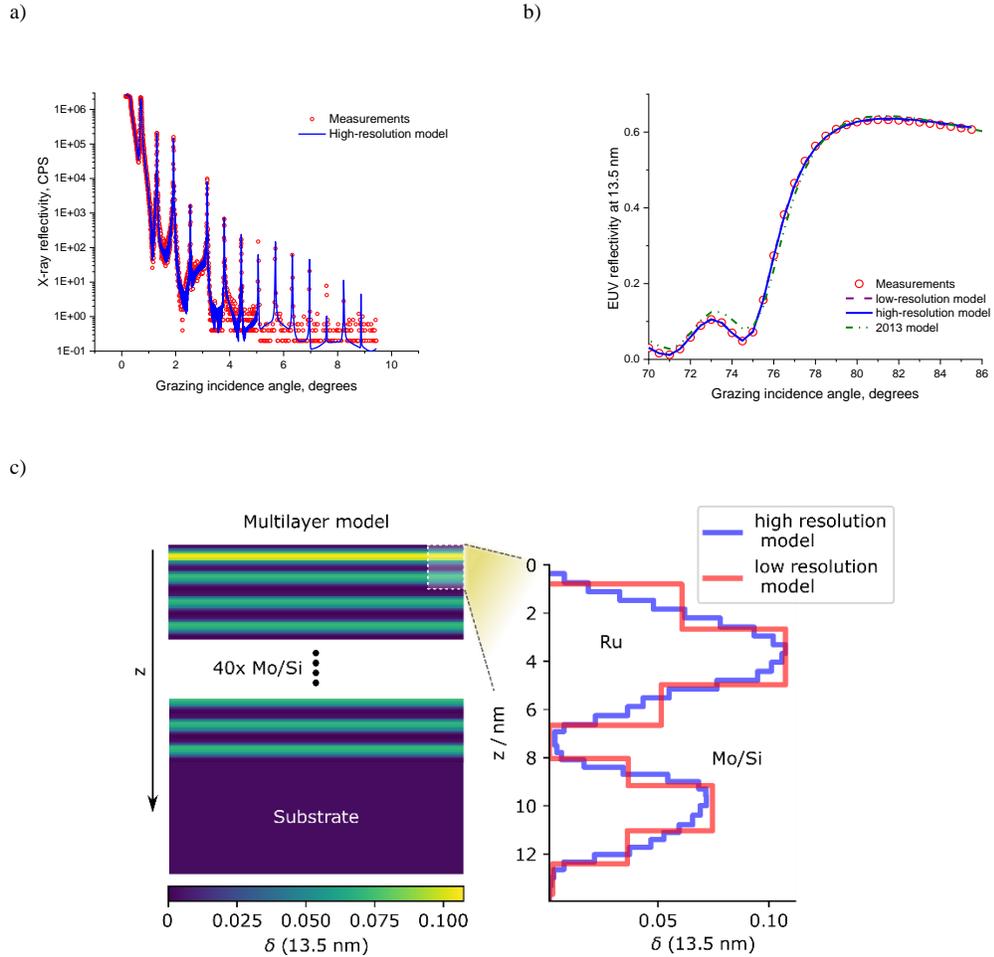

a)

b)

c)

Figure 2 (a) Measured XRR of the Mo/Si multilayer and its best fit solution. (b) Measured EUVR of the Mo/Si multilayer and the simulated EUVR from the low- and high-resolution multilayer model, as well as the simulated EUVR using the 2013 multilayer model. (c) The reconstructed low- and high-resolution δ-profiles for a single Mo/Si bi-layer calculated for EUV wavelength of 13.5nm.

The measured and simulated (using best-fit results) X-ray and EUV reflectivity curves of the MLM sample are shown in Fig. 2a and 2b, respectively. The best fitted optical constants profiles, calculated for 13.5nm wavelength are shown in Fig. 2c by δ, the decrement of the real part of optical constant *n*, *δ=real(1-n),* while the fitted complex value of of the optical constants *n*, β, can be found in Table 1.

In the δ-profile of the MLM sample (Fig. 2c) the Ru layer is shown between 0 and 7nm and the Mo/Si bilayer structure between 7nm and 14nm. We show the structure of only one Mo-Si bi-layer, because in our model we consider all 40 bi-layers to be identical. This assumption is supported by the good fit of the measured data, especially by the excellent agreement between the width of the measured and simulated Bragg peaks on the XRR curve.

The high-resolution profile shows the well-known interface asymmetry for the Mo/Si multilayer, namely the Mo-on-Si is larger than Si-on-Mo interface [10]. The low-resolution profile of the periodic MLM part just roughly describes the high-resolution shape. However, as was mentioned afore, both high- and low-resolution profiles are equally good for fitting of the EUVR measurements. Figure 2b shows that both new high- and low-resolution models fit much better to the new EUVR measurements than the previous model from 2013 [1], possibly due to the modification of the multilayer structure.

## 2.2 The absorber model

The analysis of the absorber layer was a bit more complicated, since the same approaches we used in the analysis of the multilayer mirror with tabulated CXRO atomic scattering factors did not yield a consistent fit for X-ray and EUV reflectivity, although the solution for X-ray was relatively straightforward. The reasons may be the uncertainty of the exact chemical composition and the inaccuracy of atomic scattering factors for mask materials. The manual correction of both δ and β of optical constants for -4% for TaBN and +50% for thin surface TaBO layers of the EUV optical constants were necessary to obtain the good combined agreement between EUVR and XRR fitting. This assumption about such correction of optical constants for TaBN layer was confirmed by the rigorous analysis of the optical constants for the TaBN/TaBO structure in the EUV wavelength range performed at PTB (not presented here).

The figures 3a and 3b show that the high-resolution model reasonably well fits to both X-ray and EUV reflectivity measurements. Figure 3b also shows that the low-resolution model fits even better to EUVR measurements than the high-resolution model. This is expected as the low-resolution model was obtained by fitting of EUVR-only measurements and presents one of the deep local minima, while high-resolution model can be seen as a Pareto optimal solution for the combined X-ray and EUV reflectivity data sets. The comparison between high- and low-resolution δ-profiles for 13.5nm wavelength and high-resolution profile for 0.154nm wavelength are shown in Fig. 3c.

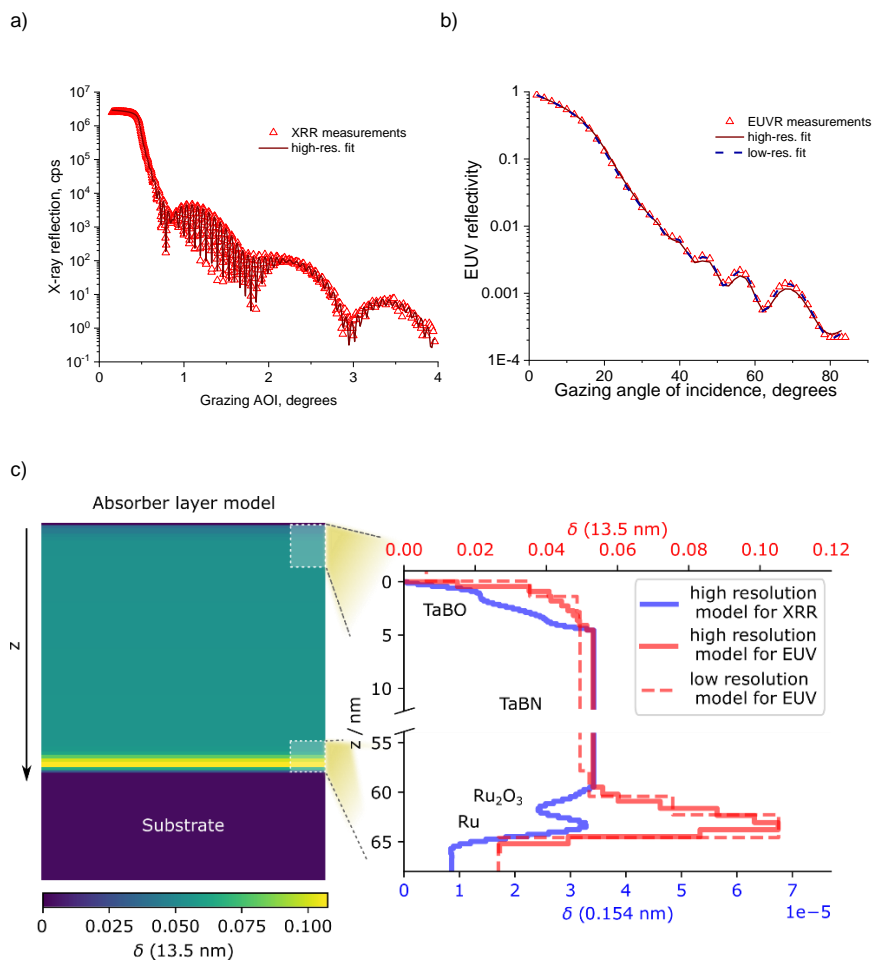

a)

b)

c)

Figure 3 Measured and best fit solutions for Ru/TaBN/TaBO thin film.(a) measured and best-fit high-resolution model calculated curves for XRR; (b) measured best-fit high- and low-resolution models for EUVR; (c) real part of decrement of the constants profiles ($\delta$) shown for 0.154nm and 13.5nm wavelengths for high-resolution model together with low-resolution $\delta$ calculated only for 13.5nm wavelength.

From the high-resolution model we can conclude that only 55nm of the nominally 60nm thick absorber layer has a constant density while the rest is consumed by interfaces and transition layers. Fig. 3c shows that the top 2nm TaBO layer most likely forms only oxygen rich interface transition regions. The underestimation of O content in TaBO model can explain the necessity for manual increase of surface density as 13.5nm light scatters much more effectively on oxygen atoms than X-rays due to proximity of O L absorption edge. The remarkable drop, visible on the $\delta$ profile calculated for X-ray wavelength at z=0nm, can be an indication of the presence of an oxidized Ru layer, what might be caused by exposure of the sample to ambient between depositions of Ru and Ta-based absorber layers. It should be noted that, as the uncertainty of EUV-only data analysis are quite large, the high- and low-resolution models can coincide within the uncertainty corridors of low-resolution profile.

For EUV lithography imaging simulations we have combined the low-resolution models of the multilayer and absorber layer stack. During the combination we fixed the Ru layer as it was determined for the MLM sample. The combined multilayer and absorber low-resolution model of the analyzed EUV mask is presented in Table 1. In the following part we compare the EUV lithography imaging simulations performed using the low-resolution model shown in Table 1 and presented in [1] in order to analyze the influence of the mask model on high-NA imaging.

**Table 1. Low-resolution model for TaBO/TaBN absorber on a Ru/Mo/Si MLM mask.**

| Layer thickness,nm | Layer optical constants for 13.5nm EUV light ($1\text{-}\delta\text{-}i\beta$) |
|---|---|
| TaBO/TaBN absorber layer | |
| 1.44 | 0.99375-0.00228i |
| 1.44 | 0.96463-0.01841i |
| 1.44 | 0.95139-0.03037i |
| 55.02 | 0.95056-0.03163i |
| 2.57 | 0.94800-0.03026i |
| Ru protective layer | |
| 1.55 | 0.93925613-0.011132225i |
| 2.465 | 0.89243499-0.016227441i |
| 1.61 | 0.94843207-0.0088824872i |
| 40x Mo/Si multilayer | |
| 1.435 | 0.99889336-0.001918392i |
| 1.215 | 0.96323059-0.0041547321i |
| 1.807 | 0.92539394-0.0063108367i |
| 1.422 | 0.96379023-0.0041027391i |
| 1.151 | 0.9976784-0.0019250528i |

## 3. Impact on key parameters in EUVL simulations

The EUV lithography simulations presented in this paper are performed with the rigorous mask 3D simulation software S-Litho EUV (Synopsys) [11]. Lines and spaces through pitch are imaged with a dipole leaf shape illumination at NA 0.55 using 4x/8x reduction system and 20% central obscuration of the projector pupil. The unpolarized EUV light is incident on the mask at 5.4 degree chief-ray angle[12]. The lines and spaces are evaluated over a pitch range from 16nm to 40nm, where the target critical dimension (CD) is the half-pitch value for pitches until 32nm and fixed to CD 16nm at larger pitches as can be seen on the curve *target CD* in Fig. 4a. The aerial image threshold is fixed to print the smallest pitch on target and the mask CD (MCD) required to print the other horizontal pitches to target at the fixed threshold is plotted in Fig. 4a. Lithography metrics such as best focus (BF), depth of focus (DoF), exposure latitude (EL) and telecentricity error (TE) are presented in Fig. 4.

The simulations denoted as "Low-res. model" are performed using the mask model presented in Table 1 and results denoted as "2013 model" are obtained with the mask model presented in [1].

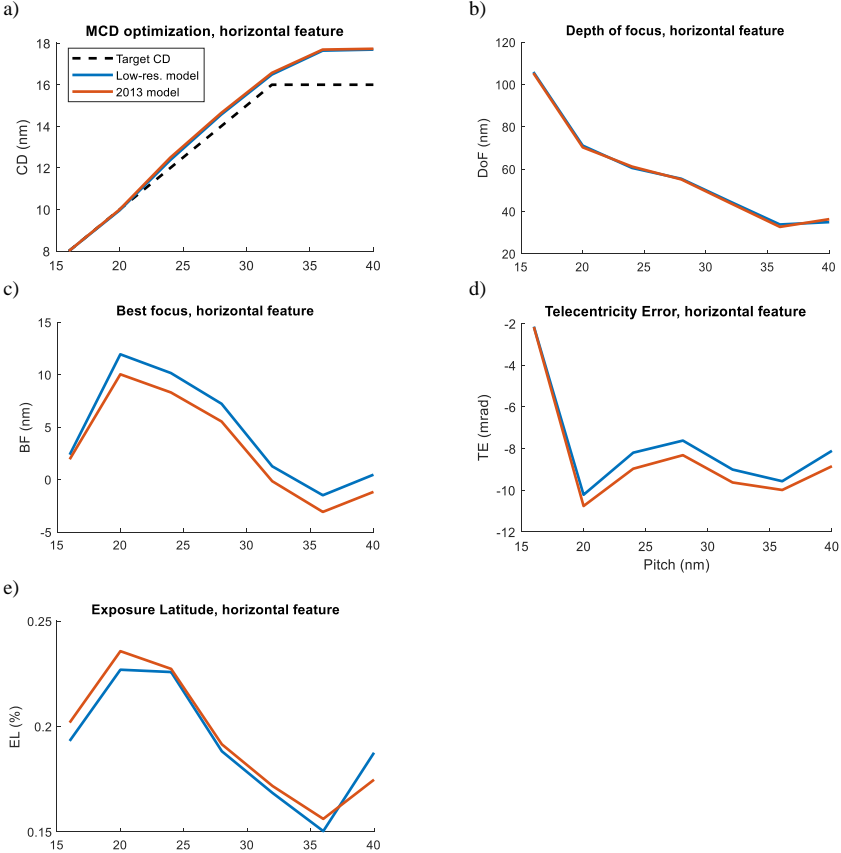

Figure 4 Comparison between EUVL metrics through pitch at high NA 0.55 calculated using the mask model presented here (denoted as Low-res. model) and presented in [1] (denoted as 2013 model).

Fig. 4 shows that regardless of the difference in the mask models there is not much difference between key EUVL metrics. The small differences can be explained by the relatively small influence of the EUVR for grazing incidence angles lower than 75 degrees (i.e., higher than 15 degrees from the normal) where EUVR curves differ most (see Fig. 2b).

## 4.  Conclusions

We have analyzed the internal structure of the current state-of-the-art EUV mask blank using XRR and EUVR measurements. The high- and low-resolution models of the optical constant profiles from the mask blank are reported. The comparison of the mask model from 2013 to the newly proposed mask model is performed for next-generation high NA EUV simulation settings on lines and spaces. The simulations show that the slight difference in measured EUVR at high incidence angles does not change the key EUVL metrics dramatically.

The high-resolution model gives an accurate description of the internal multilayer and absorber layer structures. The combined XRR and EUVR measurement of EUV mask blanks can be used to analyze changes in the multilayer and absorber structure in future mask R&D using the model described here as a reference.

Finally, we recommend the use of the new low-resolution mask model for next-generation EUVL simulations, as it fits better with the current state-of-the-art masks. The implementation of the presented low-resolution mask model should increase the simulation accuracy of more complex designs.

**Acknowledgements**

We acknowledge dr. Sergey N. Yakunin for the assistance development of free-form analysis code of combined EUV-XRR data sets and dr. Joachim Woitok and Malvern Panalytical for the help with measurement of EUV mask blanks.